\documentclass[aps, prl,twocolumn,superscriptaddress, reprint, nofootinbib, showpacs]{revtex4-1}

\usepackage{graphicx}
\usepackage{dcolumn}
\usepackage{dcolumn}
\usepackage{textgreek}
\usepackage{textcomp}
\usepackage{adjustbox}

\newcommand{\leqrange}[3]{$#1 \leq\ #2 \leq\ #3$}

\def\KtCA{K\textsubscript{2}Cr\textsubscript{3}As\textsubscript{3}}

\def\ACA{\textit{A}\textsubscript{2}Cr\textsubscript{3}As\textsubscript{3}}
\def\AoCA{\textit{A}Cr\textsubscript{3}As\textsubscript{3}}

\def\Tc{$T_c$}

\def\Na+{Na\textsuperscript{+}}
\def\Ba+{Ba\textsuperscript{2+}}

\def\tild{$\sim$}

\def\iA{\text{\AA}\textsuperscript{-1}}
\def\Asq{\text{\AA}\textsuperscript{2}}

\def\Gor{$G(r)$}
\def\Gpor{$G_p(r)$}

\def\Psmt{$P\overline{6}m2$}

\def\Ammt{$Amm2$}

\begin{document}

Notice of Copyright This manuscript has been authored by UT-Battelle, LLC under Contract No. DE-AC05-00OR22725 with the U.S. Department of Energy. The United States Government retains and the publisher, by accepting the article for publication, acknowledges that the United States Government retains a non-exclusive, paid-up, irrevocable, world-wide license to publish or reproduce the published form of this manuscript, or allow others to do so, for United States Government purposes. The Department of Energy will provide public access to these results of federally sponsored research in accordance with the DOE Public Access Plan (http://energy.gov/downloads/doe-public-access-plan).

\clearpage
\setcounter{page}{0}

\preprint{APS/123-QED}

\title{Frustrated structural instability in superconducting quasi-one-dimensional  K\textsubscript{2}Cr\textsubscript{3}As\textsubscript{3}}

\author{Keith M. Taddei}
\email[corresponding author ]{taddeikm@ornl.gov}
\affiliation{Neutron Scattering Division, Oak Ridge National Laboratory, Oak Ridge, TN 37831}
\author{Guangzong Xing}
\affiliation{Department of Physics and Astronomy, University of Missouri, Missouri 65211, USA}
\author{Jifeng Sun}
\affiliation{Department of Physics and Astronomy, University of Missouri, Missouri 65211, USA}
\author{Yuhao Fu}
\affiliation{Department of Physics and Astronomy, University of Missouri, Missouri 65211, USA}
\author{Yuwei Li}
\affiliation{Department of Physics and Astronomy, University of Missouri, Missouri 65211, USA}
\author{Qiang Zheng}
\affiliation{Materials Science and Technology Division, Oak Ridge National Laboratory, Oak Ridge, TN 37831}
\author{Athena S. Sefat}
\affiliation{Materials Science and Technology Division, Oak Ridge National Laboratory, Oak Ridge, TN 37831}
\author{David J. Singh}
\email[corresponding author ]{singhdj@missouri.edu}
\affiliation{Department of Physics and Astronomy, University of Missouri, Missouri 65211, USA}
\author{Clarina de la Cruz}
\affiliation{Neutron Scattering Division, Oak Ridge National Laboratory, Oak Ridge, TN 37831}

\date{\today}

\begin{abstract}

We present density functional theory and neutron total scattering studies on quasi-one-dimensional superconducting \KtCA\  revealing a frustrated structural instability. Our first principles calculations find a significant phonon instability, which, under energy minimization, corresponds to a frustrated orthorhombic distortion. In neutron diffraction studies we find large and temperature independent atomic displacement parameters, which result from highly localized orthorhombic distortions of the CrAs sublattice and coupled K displacements. These results suggest a more complex phase diagram than previously assumed for \KtCA\ with subtle interplays of structure, electron-phonon and magnetic interactions. 
 
\end{abstract}


\maketitle


Prior to Bardeen, Cooper and Schreiffer's theory of conventional superconductivity, several macroscopic phenomenological descriptions existed which could successfully reproduce the observed behaviors of superconductors \cite{Bardeen1957, Schafroth1954, London1935, Ginzburg1956}. It was the identification of electron-phonon coupling as the mediator of superconductivity and the subsequent microscopic theory that propelled the field forward and afforded new predictive powers \cite{Ginzburg1997, Ashcroft1968}. No such universal interaction has been identified in the study of  unconventional superconductors (UNSC) \cite{Stewart2017,Norman2011}. Yet in most UNSC, clear candidate interactions are manifested through a quasi-universal proximity to ordering instabilities - such as magnetism, charge-ordering, structural distortions or, as is often the case, confounding combinations of these orders \cite{Basov2011,Avci2011,Chubukov2012,Fernandes2014,White2015}. For the most widely studied families (e.g. cuprates, iron-based, heavy-Fermion), a vital tool for discerning the relevance of these interactions to superconductivity has been the ability to tune through the instabilities via parameters such as pressure, charge-doping and strain \cite{Chubukov2012,Monthoux2007,Lederer2017}. 

Recently, a new family of possible UNSC was discovered, \ACA\ (with \textit{A} = Na, K, Rb, Cs), which has generated significant interest due to potentially exotic physics and provocative properties such as structural quasi-one-dimensionality (Q1D), non-centrosymmetry, Luttinger-Liquid-like behavior and the transition-metal pnictide composition \cite{Bao2015, Alemany2015,Jiang2015, Wang2016, Liu2016, Zhou2017, Watson2017, Mu2018}. The main structural motif of CrAs double-walled sub-nano tubes is a fascinating contrast to the 2D transition-metal planes of the cuprates and iron-based superconductors. However, unlike its 2D brethren, the \ACA\ materials have proven difficult to tune, being unreceptive to chemical doping and robust against internal/external pressure, which only suppresses the superconducting transition temperature (\Tc )  \cite{Cao2017, Wang2016}. 


Furthermore, while measurements of dynamic phenomena have revealed signatures of spin-fluctuations, no long-range order analogous to the spin/charge density waves of the classic UNSC has been observed \cite{Taddei2017t, Zhi2015, Yang2015, Zhi2016}. Initial findings of a possible spin-glass phase in the similar 1:3:3 stoichiometry (\AoCA ) have given way to findings of robust superconductivity with no static magnetism \cite{Cao2015, Mu2017, Liu2018}. In the absence of the often-observed domed phase diagram or clear evidence of a nearby instability, the superconducting mechanism remains elusive and conflicting suggestions have developed ranging from spin-fluctuations to conventional phonons \cite{Subedi2015, Taddei2017t, Wu2015, Zhang2016}. Therefore, any evidence of instabilities could be vital to understanding this material and how it fits into the wider narrative of UNSC. 

To better understand superconductivity in \KtCA\ we performed first principles calculations, which predicted a significant in-plane phonon instability with a large degeneracy resulting in a frustrated instability of the \Psmt\ structure. As a first approach for experimental verification, we conducted a careful analysis of the atomic displacement parameters (ADPs) observing an anomalous temperature dependence which is localized to the \textit{ab} plane. Using pair distribution function (PDF) techniques to probe the local structure, we found an orthorhombic distortion localized to single CrAs tubes. These results indicate that, in addition to the previously reported proximate magnetic instability, \KtCA\ has a structural instability albeit frustrated and thus an important similarity to the wider family of UNSC \cite{Taddei2017t}.  

Density functional theory (DFT) calculations were performed using the general gradient approximation, projector-augmented wave method and the general potential linearized augmented plane-wave method with the experimental lattice parameters (see supplemental material [\onlinecite{SM}])\cite{Perdew1996, Kresse1999, Kresse1996, Singh2006, Togo2015}. Neutron total scattering measurements were performed on the NOMAD time-of-flight diffractometer of Oak Ridge National Laboratory's Spallation Neutron Source and the data reduced using the instrument developed software \cite{McDonnell2017,Neuefeind2012}. The diffraction data were used (along with the data reported in Ref.~[\onlinecite{Taddei2017t}]) for Rietveld refinements (as implemented in the GSAS and EXPGUI software suites) to model the ADPs and their temperature dependence \cite{Toby2001, Larson2004}. For our PDF analysis, least-squares refinements were performed as implemented in PDFgui \cite{Farrow2007, Proffen1999}. 



\begin{figure}
	\includegraphics[width=\columnwidth]{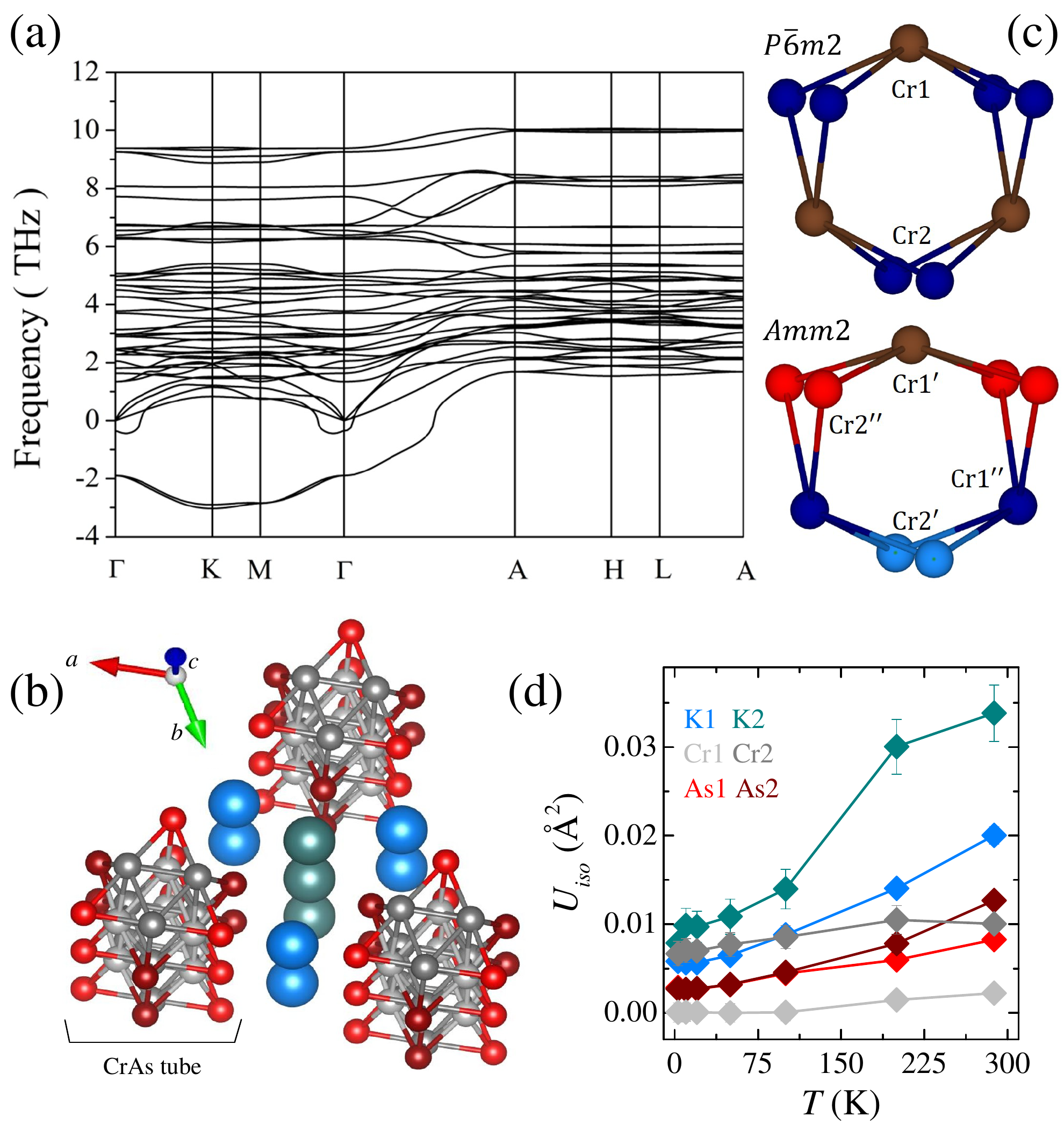}
	\caption{\label{fig:one}  Results of Density Functional Theory (DFT) calculations and preliminary diffraction analysis. (a) Phonon dispersions for selected high-symmetry directions through the Brillouin zone. (b) Structural model emphasizing the CrAs tubes and the three-fold $C_{3}$ symmetry. (c) Initial \Psmt\ Cr sublattice used as the input structure for DFT and the predicted \Ammt\ structure found in structural relaxation under energy minimization. (d) Temperature dependence of the atomic displacement parameters (U), from Rietveld refinements performed with diffraction data. The atom labels and color coding used in panel (d) and (b) are consistent though not with (c) where more colors were needed due to the number of sites. }	
\end{figure}

Initially motivated to check for a potential conventional superconducting mechanism, we performed DFT calculations to study electron-phonon coupling in \KtCA . Unexpectedly, our calculated phonon spectrum (Fig.~\ref{fig:one}(a)) predicted a substantial structural instability of the CrAs tubes - in contrast to previous work \cite{Subedi2015}. This is apparent in the negative energy optical phonon branch observed along the in-plane $\Gamma  - K, K - M$ and $ M - \Gamma $ directions (standard labels are used as in Refs.~[\onlinecite{Alemany2015}] and [\onlinecite{Jiang2015}]). The existence of a negative energy phonon mode suggests a distortion to a lower symmetry structure. However, the unstable branches are nearly dispersionless in-plane which should confound the selection of a minimum energy distortion direction. By comparison, significant dispersion is seen along the tube direction ($\Gamma - A$). We interpret this in-plane phonon energy degeneracy as representing a frustration that would work against long-range ordering in-plane. 

To characterize the possible structural distortion, we allowed the crystallographic parameters to relax following energy minimization. We started from the reported \Psmt\ structure (Fig.~\ref{fig:one}(b)) which consists of two in-equivalent alternating stacked planes at $z = 0, \frac{1}{2} $. Each plane has unique K, Cr and As sites with atomic labels and Wyckoff positions: (K1($3k$), Cr1($3k$), As2($3k$)) and (K2($1c$), Cr2($3j$), As1($3j$)) for $z = 0, \frac{1}{2}$ respectively. Under energy minimization we find the structure undergoes a $\Gamma$ point distortion to the \Ammt\ orthorhombic symmetry due to the unstable phonon mode (Fig.~\ref{fig:one}(c)). This breaks the $\overline{6}$ roto-inversion symmetry and splits each of the Cr1(2), As1(2) and K1 sites into two. Additionally, broken mirror symmetries allow for displacements along the (210) direction. 

To check for possible evidence of this distortion in the measured properties, we calculated the density of states for both the \Psmt\ and \Ammt\ structures. While the expected frustration of this order complicates the interpretation, we do not observe large changes which would predict significant modifications to transport properties (see SM). Therefore, we rely on diffraction studies to search for the predicted distortion. Previously, we reported the \Psmt\ model as adequate for fits of high resolution diffraction data \cite{Taddei2017t}. This limits physical manifestations of the DFT results, indicating either a small (resolution limited) upper limit to the distortion magnitude or a local distortion model. The latter agrees with our predicted lattice frustration encouraging a study of the local structure. Therefore, we used neutron diffraction data collected to high scattering angle to model the isotropic ADPs ($U_{iso}$) via standard Rietveld refinement techniques in the \Psmt\ structure. ADPs measure the mean square variance in atomic positions and can give evidence of lattice disorder via large values or temperature independent behavior \cite{Cruickshank1956,Dunitz1997,Sales1999,Sales2001}.  

The resulting temperature dependent ADPs are shown in (Fig.~\ref{fig:one}(d)). Comparing our values to related materials we find ADPs significantly larger than in the structurally related (and stable) Q1D Chevrel phases and higher than those reported in the iron-based superconductors, which are famous for their local nematicity \cite{Gall2014,Villars2016,Frandsen2017}. Moreover, the values obtained for the K site are similar to the Einstein \lq rattler\rq\ behavior seen in thermoelectric materials \cite{Sales2001, Sales2000}. To this latter comparison, we similarly observe significant residual $U_{iso}$ at 2 K on the K2 site. The Cr2 site not only exhibits a relatively large residual ADP but also shows little temperature dependence over the measured range. In contrast, the As1, As2 and Cr2 sites show Debye-Waller type behavior, tending towards low values as $T\rightarrow 0$ K. The temperature dependence of the former three sites, indicates ADPs not exclusively driven by standard thermal effects.  

\begin{figure}
	\includegraphics[width=\columnwidth]{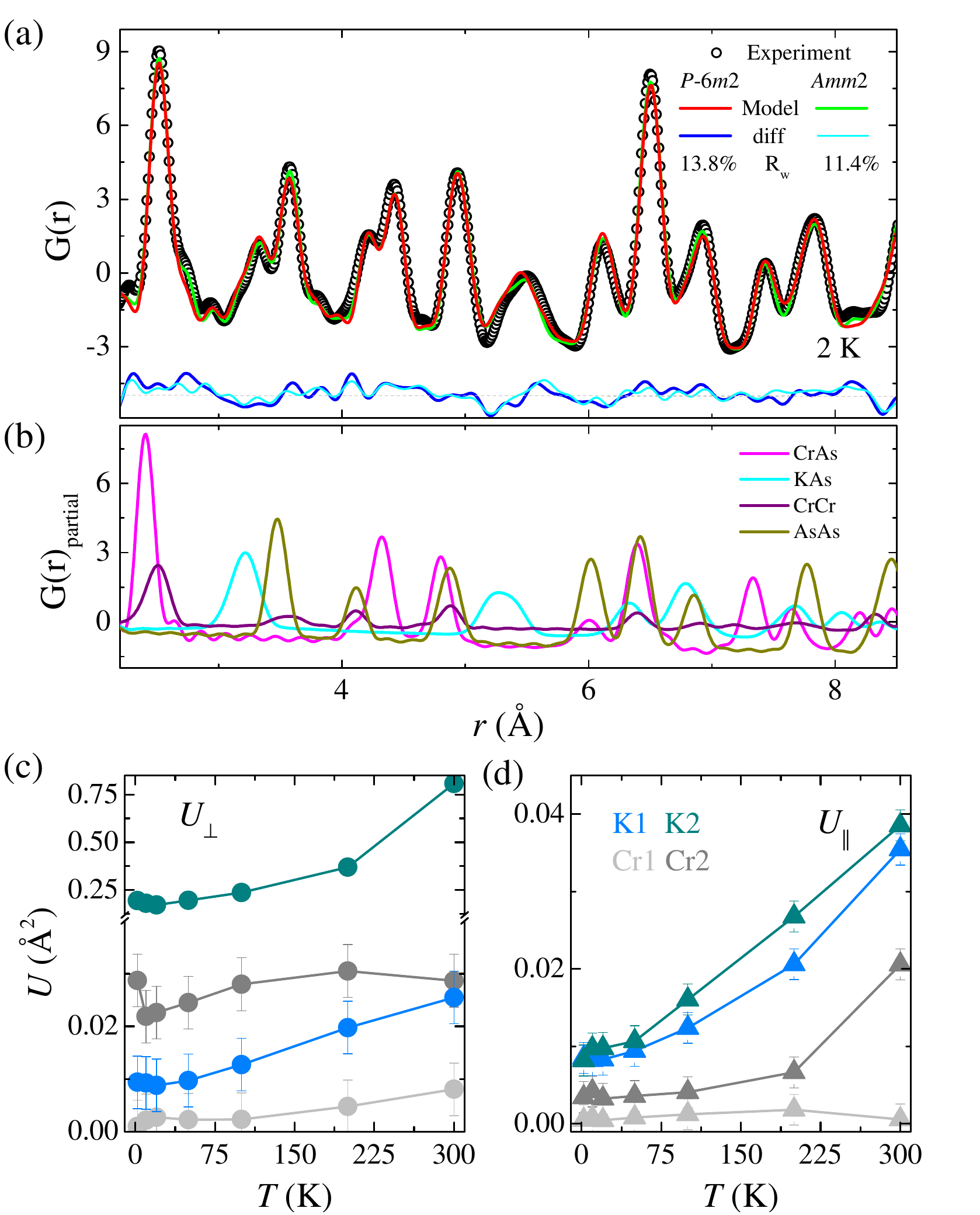}
	\caption{\label{fig:two} Results of pair distribution function (PDF) analyses. (a) Fit using $G(r)$ data (where $r$ is a real-space distance measured in \AA ) using the averaged $P\overline{6}m2$ and distorted $Amm2$ structure symmetry constraints. (b) Partial PDF showing the K-As, Cr-As, Cr-Cr and As-As correlation functions for a $P\overline{6}m2$ fit. Anisotropic \textit{U} parameter determined from PDF analysis for (c) \textit{ab} plane ($U_{\perp}$) and (d) lattice \textit{c} direction ($U_{\parallel}$). ADPs for As1(2), which show unremarkable temperature dependence, are not shown in panels (c) and (d) to provide visual clarity.}	
\end{figure}


To better characterize the structural effects causing the large ADPs, we performed PDF analysis, which takes the Fourier transform of the normalized diffraction pattern to generate the real space atomic correlation function \Gor . Fitting the transformed data shifts the features of the fit to direct atom-atom spacings (measured by $r$ in \AA ), rather than crystallographic symmetry. Therefore, it allows for the modeling of the local structure through least-squares techniques \cite{Farrow2007, Proffen1999}.    

Figure~\ref{fig:two}(a) shows \Gor\ for \KtCA\ collected at 2 K and a corresponding fit using the symmetry constraints of the \Psmt\ structure. Here, we model only the low-\textit{r} region of 2.1 - 20 \AA , which covers approximately the first two unit cells of the hexagonal structure and therefore accentuate the local structure in the fit (see SM). We start with the symmetry constraints of \Psmt\ to refine our discussion of the ADPs, which are more directly modeled in \Gor\ as peak widths. Motivated by the co-planar arrangement of the atomic sites, we treat the ADPs anisotropically with in-plane ($U_{\perp}$) and out-of-plane ($U_{\parallel}$) components (Fig.~\ref{fig:two}(c) and (d)). This analysis reveals that the large and temperature independent ADPs of K2 and Cr2 shown in Fig.~\ref{fig:one}(d) arise from $U_{\perp}$ which displays both behaviors. Meanwhile, $U_{\parallel}$ is mostly Debye-Waller like (Fig.~\ref{fig:two}(d)).  The $U_{\perp}(2 \text{ K})$ values are quite large compared to similar materials and indicate possible local disorder of the K2 and Cr2 sites within the \textit{ab}-plane as expected for an \Ammt\ type distortion \cite{Gall2014,Villars2016}.   

Beyond large ADPs, we note the generally poor fit of the \Psmt\ model to the low-\textit{r} PDF data (Fig.~\ref{fig:two}(a)). Several features are missed, and intensities mismatched which result in a relatively large fit residual $R_w$ (for reference a crystalline standard results in $R_w \sim\ 10 \%$). Despite the complicated CrAs tube geometry with many similar (overlapping in \Gor) bond lengths (Fig.~\ref{fig:two}(b)), a series of peaks ($r = 2.6 \text{ and } 5.4$ \AA ) can be identified which are missed by the \Psmt\ model (Fig.~\ref{fig:two}(a)). This suggests additional atom-atom distances not allowed in the average structure. The anomalously large ADPs of K2 and Cr2 result from the model attempting to fit these extra correlation lengths - broadening the peaks and reducing the peak intensity to account for the smaller coordination number. If we model with ADP values of similar materials and structures (i.e. with smaller ADPs), the intensity mismatch is exacerbated, supporting this interpretation \cite{Villars2016, Gall2014}. 


To account for these features, we attempted modeling of our PDF data allowing for atomic displacements consistent with the maximal symmetry subgroups of \Psmt\ - including \Ammt . Of these, we focus on the \Ammt\ model as the highest symmetry distortion capable of improving our PDF  fit (see SM). As in the \Psmt\ refinements, at first we refined the atomic sites with the symmetry constraints of the space group. Additional constraints were applied to ensure the same number of refinable parameters in each model - an important condition to prevent over-parameterization of the limited $r$ range considered. 

As discussed, the \Ammt\ model splits the $3k$ and $3j$ Wyckoff sites into two independent sites each and thus breaks the degeneracy of several bond lengths. In doing so, the \Ammt\ structure can model the previously missed peaks and reduces the $R_w$ parameter from 13.8\% to 11.4\% (Fig.~\ref{fig:two}(a)). Furthermore, these newly allowed displacements reduce the $U$ for K2 from $0.30$ to $0.03$ \Asq\ (for first unit cell $r$-range). This agrees with our suggestion of the large ADPs as an attempt to fit the split peaks.

\begin{figure}
	\includegraphics[width=\columnwidth]{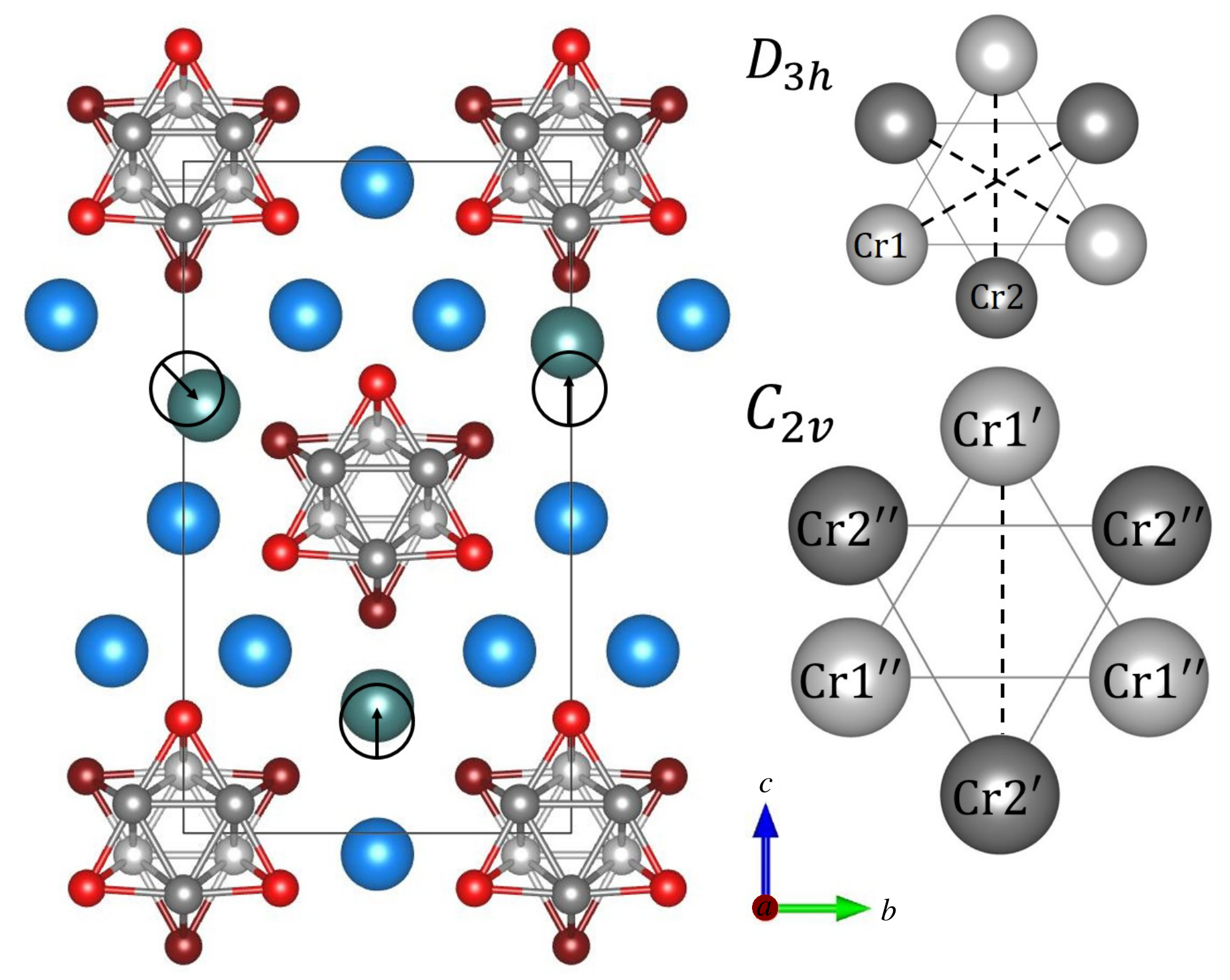}
	\caption{\label{fig:three}  (left) \Ammt\ structure determined in low-\textit{r} PDF fits, the results of the the three site K2 refinement are shown, with black circles and arrows marking the \Psmt\ K2 position and the direction of the displacement respectively. This reveals a lack of displacement direction correlation between neighboring K2's. (right) Symmetry of the Cr tubes in the \Psmt\ and \Ammt\ space groups (point groups $D_{3h}$ and $C_{2v}$ respectively) with mirror planes represented by dashed lines.}	
\end{figure}


The fit \Ammt\ structure is shown in Fig.~\ref{fig:three}. The main differences from \Psmt\ (Fig.~\ref{fig:one}(b)) are a large displacement of the K2 site along the orthorhombic (210) direction and the breaking of the CrAs tubes' $C_3$ rotational symmetry. Previous DFT work has suggested the K count and Cr-Cr bonding coupled through the electron distribution \cite{Alemany2015,Wu2015,Zhi2015}. Due to the bonding(antibonding) nature of the Cr-Cr(Cr-As) bonds, the displacement of K2 towards the CrAs tube should decrease(increase) the respective bond lengths or \textit{vice versa} \cite{Alemany2015}. This is what we see, as the triply degenerate 2.66 \AA\ Cr2-Cr2 bond breaks into a 2.55 \AA\ and two 2.65 \AA\ bonds and the 2.49 \AA\ Cr2-As1 bond breaks into a 2.58 \AA\ and two 2.51 \AA\ bonds - the shorter Cr-Cr and longer Cr-As corresponding to the Cr closest to the displaced K (Fig.~\ref{fig:three}).  

\begin{figure}
	\includegraphics[width=\columnwidth]{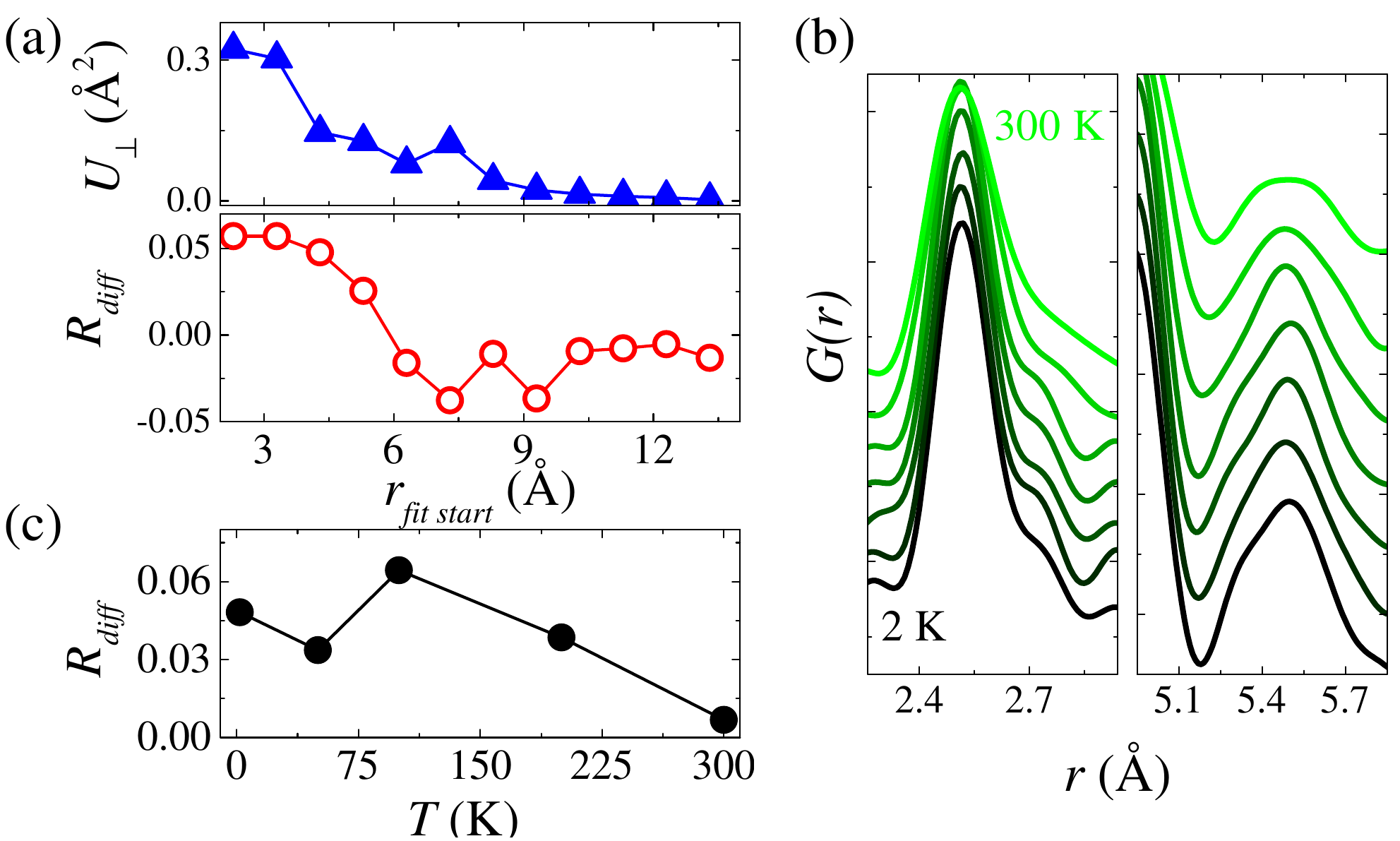}
	\caption{\label{fig:four}  (a) $U_{\perp}$ and the normalized \textit{R}-factors ($R_{diff}$) as functions of starting $r$ ($r_{fitstart}$) for boxcar PDF fits using \Ammt\ and \Psmt\ symmetry constraints . (b) Waterfall plots showing temperature dependence of the \Gor\ 2.7 and 5.4 \AA\ peaks. Patterns collected at 2(black), 10, 20, 50, 150, 200 and 300 K (green). (c) Temperature dependence of $R_{diff}$ for low-$r$ PDF fits.}	
\end{figure}

The analysis of the standard diffraction and PDF data indicate significant local distortions that are uncorrelated over long distances, averaging to \Psmt . To investigate the scale of this transition, boxcar fits of the PDF data were performed by choosing a window size ($\Delta r \sim\ 9$ \AA) and then sliding the starting $r$ value ($r_{fit start}$) of the fit range to increasing \textit{r}. In this way, the analysis can be shifted from the local structure to the approximate averaged structure (high \Gor\ approximates the averaged structure as larger coordination distances are invoked) and the relative fit residuals of the different models can be investigated over different correlation lengths. 


The normalized difference of $R_w$ ($R_{diff} = 2 * (R_{orth} - R_{hex})/( R_{orth} + R_{hex})$) from boxcar fits is shown in Fig.~\ref{fig:four}(a). For the lowest $r_{fit start}$, the orthorhombic model results in significantly improved $R_{diff}$. However, by $r_{fit start} = 6.3$ \AA\ the hexagonal model produces nearly identical $R_w$ to the orthorhombic model - indicating that by 6.3 \AA\ modeling with ADPs is as good as with the orthorhombic distortion. At higher $r_{fit start}$, $R_{diff}$ settles to zero. Similarly, we can fit $U_{\perp}$ as a function of $r_{fit start}$, (Fig.~\ref{fig:four}(a)) revealing a quick drop-off of the K2 ADP which approaches the value found in standard diffraction analysis ($U_{\perp} \approx\ 0.01$ \AA $^2$)  by $r_{fit start} = 9$ \AA . 

These distance scales are approximately the size of the CrAs tubes but smaller than the intertube spacing ($\sim$10 \AA ) indicating the distortion is essentially uncorrelated between unit cells. We note that each CrAs tube has $C_3$ rotational symmetry  and therefore can distort along any of three directions. When $r_{fit start} = 6.3$ \AA\ correlations over 15 \AA\ are considered including the next nearest CrAs tube and 8 in-plane K2 correlations for every CrAs tube and the orthorhombic distortion averages to the \Psmt\ structure. This is consistent with inherent frustration indicated by the degeneracy of the relevant phonon mode in the $k_z=0$ plane. In fact, models using $Pmm2$ symmetry which creates two distinct CrAs tubes, further reduce $R_w$.  However, $Pmm2$ produces more refinable parameters and direct comparisons of $R_w$ with the \Psmt\ model are more ambiguous, we therefore base our conclusions on the \Ammt\ distortion. 

To corroborate this length scale, we performed PDF refinements with the symmetry constraints of the K2 sites in the \Ammt\ model removed. This introduces fewer additional parameters than lower symmetry space groups while still allowing us to check the independence of the tubes. This model resulted in uncorrelated displacements between the K2 sites (Fig.~\ref{fig:three}) which nonetheless remained along the mirror planes of the $C_3$ symmetry. We suggest the distortion happens randomly from CrAs to CrAs tube, such disorder would give rise to the observed average structure and a large in-plane ADP for the K2 and Cr2 sites over long correlation lengths.  

Identifying whether these distortions are static or dynamic is of great interest to their relevance to superconductivity. However, the technique of neutron total scattering integrates over  finite energy transfer and the resulting PDF pattern includes dynamic effects with the energy scale of several meV (or $10^{-13}$ s) \cite{Frandsen2017}. The temperature dependence of the distortion can instead be used as an indirect probe for the time-scale of the correlations, with temperature independence being expected of static effects.  Figure~\ref{fig:four}(b) shows the two clearest split peaks between 2 and 300 K. In both cases the distortion is seen to persist to 300 K - albeit with considerable broadening. We also performed fit comparisons between the two models over the measured temperature range (Fig.~\ref{fig:four}(c)), and found the orthorhombic model to always produce reduced $R_w$. This temperature dependence is more indicative of a static distortion than of fluctuations. However, further studies of the lattice dynamics are needed to clarify this important distinction. 

     
Our results indicate that \KtCA\ is susceptible to an orthorhombic distortion which is frustrated by a nearly degenerate in-plane dispersion of the relevant phonon mode. We find distortions of the CrAs tubes and coupled K site displacements consistent with a DFT predicted \Ammt\ symmetry lowering. However, no significant correlation from unit cell to unit cell occurs preventing long range orthorhombic order. That \KtCA\ has such an instability is significant to the material's position within the wider family of UNSC. Such a lattice distortion more clearly places it on similar footing with the cuprate and iron-based materials which tend to exhibit structural and magnetic instabilities. 

\begin{acknowledgments}
The part of the research that was conducted at ORNL’s High Flux Isotope Reactor and Spallation Neutron Source was sponsored by the Scientific User Facilities Division, Office of Basic Energy Sciences, U.S. Department of Energy. The research is partly supported by the U.S. Department of Energy (DOE), Office of Science, Basic Energy Sciences (BES), Materials Science and Engineering Division. The authors thank B.A. Frandsen, M. McDonnell and  T.-M. Usher-Ditzian for guidance with PDFgui and J. Neuefeind for help with data collection on NOMAD.
\end{acknowledgments}

\end{document}


\preprint{APS/123-QED}

\title{Supplemental Material for: \\ Frustrated structural instability in superconducting quasi-one-dimensional  K\textsubscript{2}Cr\textsubscript{3}As\textsubscript{3}}

\author{Keith M. Taddei}
\email[corresponding author ]{taddeikm@ornl.gov}
\affiliation{Neutron Scattering Division, Oak Ridge National Laboratory, Oak Ridge, TN 37831}
\author{Guangzong Xing}
\affiliation{Department of Physics and Astronomy, University of Missouri, Missouri 65211, USA}
\author{Jifeng Sun}
\affiliation{Department of Physics and Astronomy, University of Missouri, Missouri 65211, USA}
\author{Yuhao Fu}
\affiliation{Department of Physics and Astronomy, University of Missouri, Missouri 65211, USA}
\author{Yuwei Li}
\affiliation{Department of Physics and Astronomy, University of Missouri, Missouri 65211, USA}
\author{Qiang Zheng}
\affiliation{Materials Science and Technology Division, Oak Ridge National Laboratory, Oak Ridge, TN 37831}
\author{Athena S. Sefat}
\affiliation{Materials Science and Technology Division, Oak Ridge National Laboratory, Oak Ridge, TN 37831}
\author{David J. Singh}
\email[corresponding author ]{singhdj@missouri.edu}
\affiliation{Department of Physics and Astronomy, University of Missouri, Missouri 65211, USA}
\author{Clarina de la Cruz}
\affiliation{Neutron Scattering Division, Oak Ridge National Laboratory, Oak Ridge, TN 37831}

\date{\today}

\maketitle

\section{\label{sec:exp} Experiment Methodology}

\KtCA\ was synthesized as described in Ref.~\onlinecite{Taddei2017t}. For total scattering experiments on the NOMAD diffractometer $\sim 2$ g of material was loaded into a 6 mm diameter vanadium PACS can. All sample handling was performed in a high purity He glovebox to protect the sample from exposure to air. Measurements were carried out in a crysostat over the temperature range $ 2 \text{K} \leq\ T \leq\ 300 \text{K}$. Data were collected at 2, 50, 150, 200 and 300 K and in 2 K steps between 2 and 20 K with collection times of 4 hrs and 30 mins respectively. Scans of an empty 6 mm vanadium PACS can were collected in a cryostat for background correction with 3 hrs of count time. All scans were performed on warming.

\begin{figure}[h]
	\includegraphics[scale=0.5]{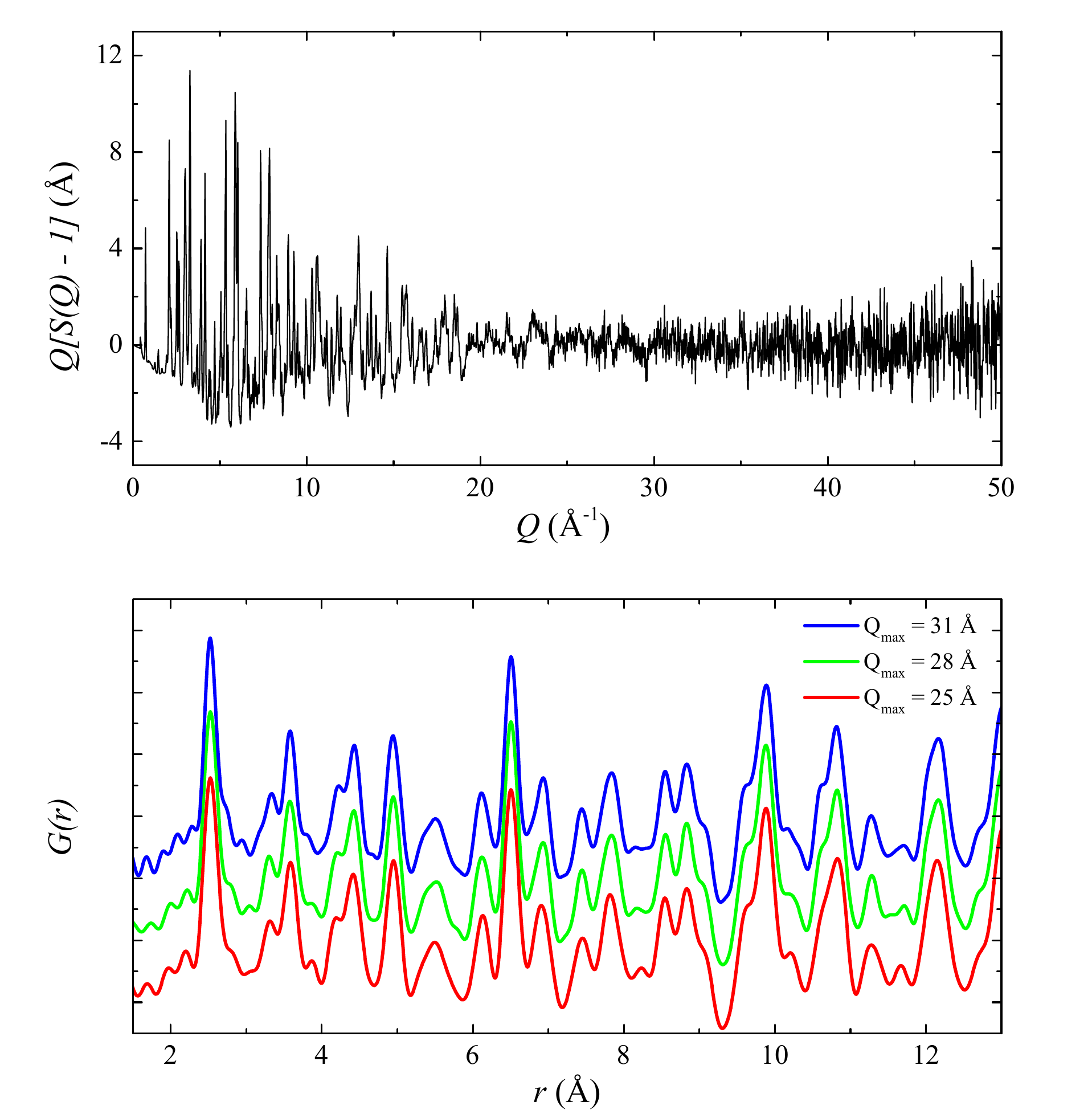}	
	\caption{\label{fig:sone} Normalized and $Q$ scaled total scattering structure factor $Q[S(Q)-1]$ (top). Radial distribution function \Gor\ for $Q_{max} = $ 25, 28, and 31 \iA\ showing artifacts due to termination ripples and noise (bottom).}	
\end{figure}

The resulting diffraction patterns were transformed into \Gor\ using the fastgr software \cite{McDonnell2017}. It is an inescapable limitation of physical diffractometers that only a finite range of usable $Q$'s can be collected in part due to: limitations of scattering geometry, finite sample size and scattering properties of the sample. In the Fourier transform of $S(Q)$ to \Gor\ this results two completing concerns when deciding were to truncate the transform (e.g. at what $Q_{max}$). The larger the range of $Q$ used in the transformation the higher the resolution of the resulting $G(r)$ however, as higher $Q$ is including the strength of the sample scattering decreases and noise becomes more prominent approaching and surpassing the size of the sample signal (Fig.~\ref{fig:sone}) \cite{Proffen1999}. This noise present at high $Q$ when transformed leads to large artificial ripples in \Gor\ which can contaminate analysis \cite{Proffen1999}. 

In order to check for such artifacts, we performed data reduction with varying $Q_{max}$ values. The resulting \Gor\ for $Q_{max} = $ 25, 28, and 31 \iA\ are shown in Fig.~\ref{fig:sone} with an arbitrary vertical offset. One way to characterize these artifacts is via the magnitude, width and peak position dependence on $Q_{max}$ of the low-\textit{r} oscillations seen below the lowest known bond (i.e. $\sim 2.3$ \AA). In our patterns these features are smaller than our considered signal and show a significant position dependence on $Q_{max}$. As a further check, analysis of fits using the \Psmt\ and \Ammt\ models discussed in the main text were performed for the three \Gor 's and, in each fit, produced similar results. 

For our final PDF analysis, a $Q_{max}$ of 31 \iA\ was used for $T < 200$ K. For higher temperature data, where larger thermal vibrations lead to a quicker drop off in the structure factor, a $Q_{max}$ of 28 \iA\ was used. 

\newpage

\section{\label{sec:GSAS} $I(Q)$ Refinements}

\begin{figure}[h]
	\includegraphics[scale=0.5]{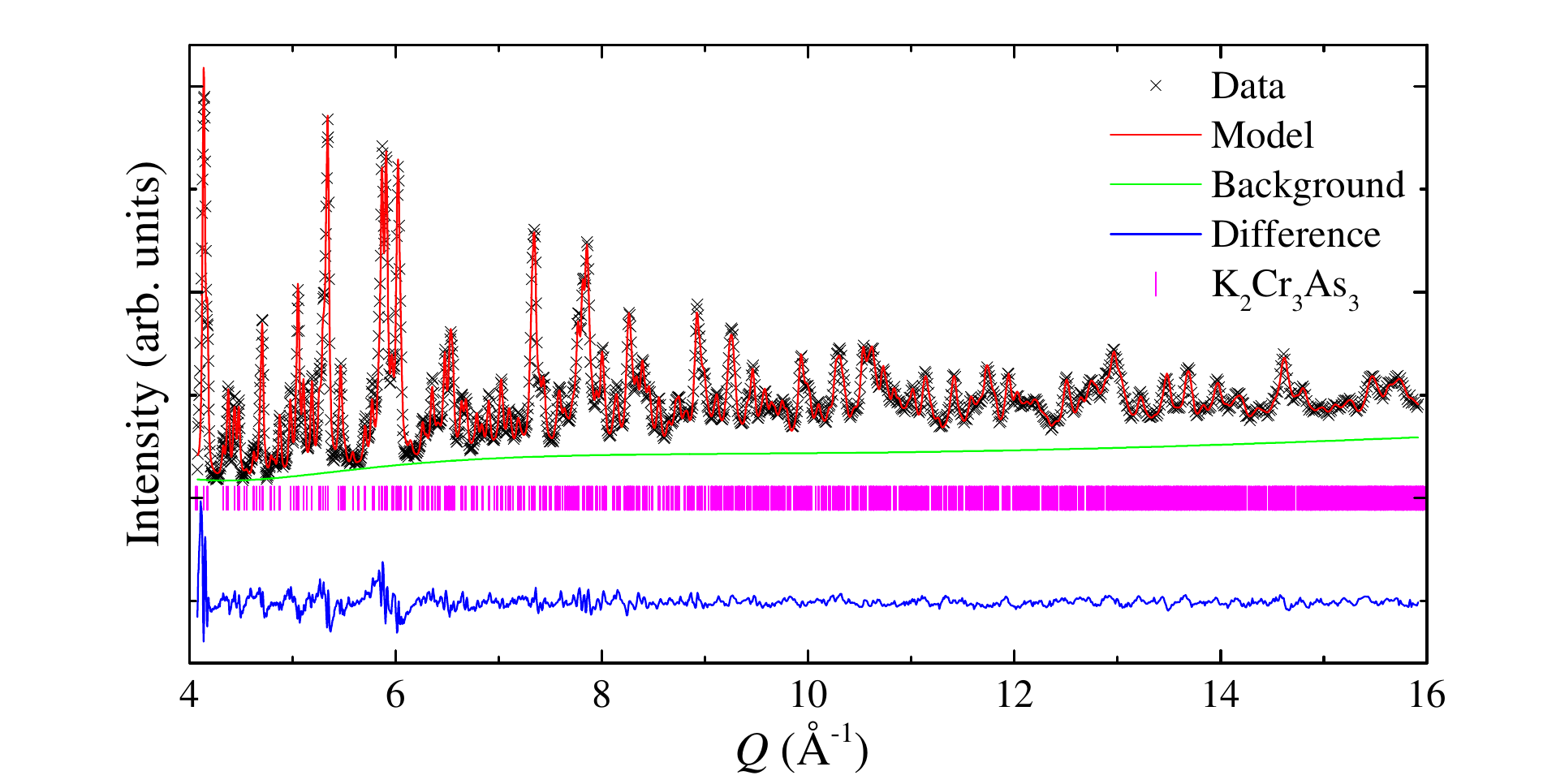}	
	\caption{\label{fig:stwo} Rietveld fit of \Psmt\ model using the 2 K data collected by BANK 5 of the NOMAD diffractometer}	
\end{figure}

Rietveld refinements using the standard diffraction data technique were performed using the GSAS and EXPGUI software suites \cite{Toby2001,Larson2004}. The time-of-flight profile function TYPE-3 of GSAS was used to model the peak profile via convoluted back-to-back exponentials convoluted with a psuedo-Voigt \cite{VonDreele1982}. All data was well fit by a single phase using the average \Psmt\ space group of \KtCA . A fit to the 2 K data is shown in Fig.~\ref{fig:stwo} with goodness of fit parameters $\chi ^{2} = 13.88, R_p = 6.7\% \text{ and } R_{wp} = 6.7\%$. Similar fits were obtained for all temperatures.   

\newpage

\section{\label{PDF} Pair Distribution Function Refinements}

\begin{figure}[h]
	\includegraphics[scale=0.5]{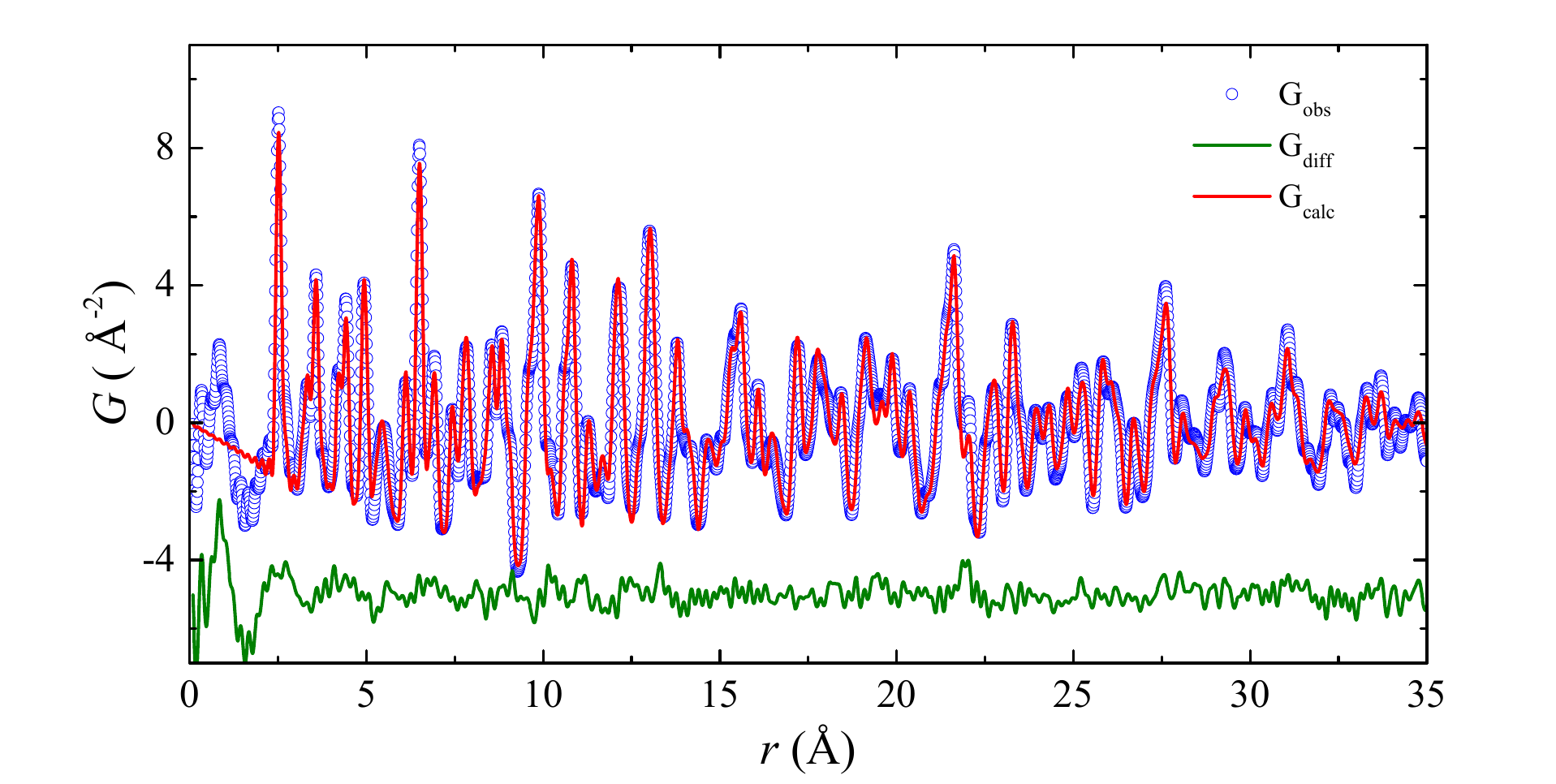}	
	\caption{\label{fig:sthree} PDFgui fit to \Gor\ collected on NOMAD at 2 K using the \Psmt\ model. }	
\end{figure}

Small box modeling of the pair distribution function \Gor\ was performed using the PDFgui software \cite{Farrow2007,Proffen1999}. The instrumental terms $Q_{damp}$ and $Q_{broad}$ were determined by refinements using data collected on the instrument standards (silicone and diamond powders) in an Orange crystostat and fixed for all modeling of the \KtCA\ data. Figure.~\ref{fig:sthree} shows a best fit using the \Psmt\ symmetry of the 2 K Fourier transformed NOMAD data over the range \leqrange{0.1 \text{ \AA}}{r}{35 \text{ \AA}} with fit parameters of $\chi ^2$ = 0.1 and $R_w$ = 16\% . As seen the model has difficulties reproducing both the low and high \textit{r} data leading to the poor fit parameters (we note that the goodness of fit parameters for the standards in the cryostat were $\chi ^2 \sim$ 1.5 and $R_w \sim$ 11\% (for the same \textit{r}-range) indicating the baseline for quality of fit).  

\begin{figure}[h]
	\includegraphics[scale=0.5]{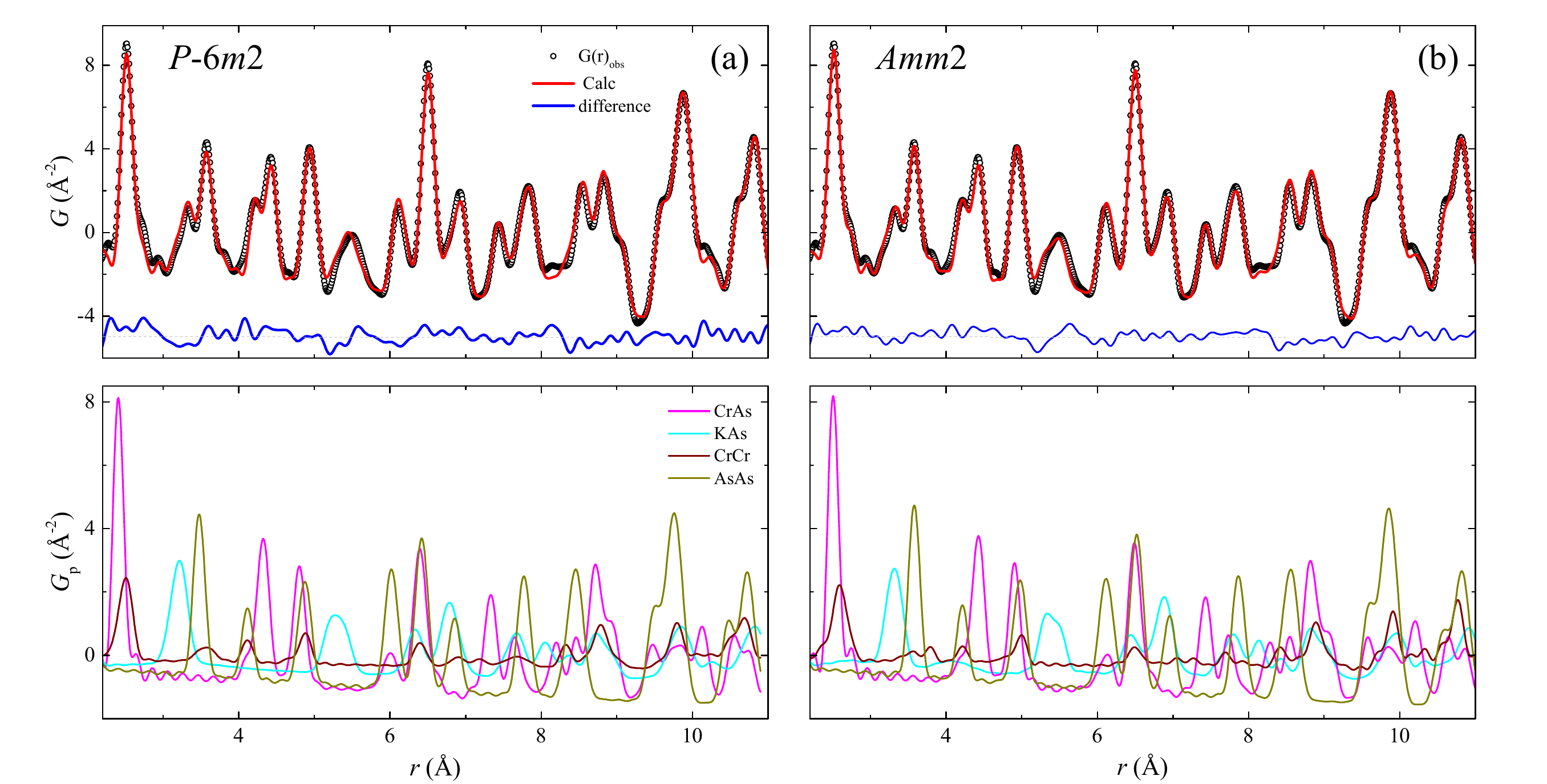}	
	\caption{\label{fig:sfour} PDFgui fits to 2 K \Gor\ patterns comparing the \Psmt\ (a) and \Ammt\ (b) models. The lower panels show the partial \Gor 's for the Cr-As, K-As, Cr-Cr and As-As correlations.  }	
\end{figure}

Figure~\ref{fig:sfour} shows a comparison of the low-\textit{r} PDFgui fits using the \Psmt\ and \Ammt\ models and the constituent partial \Gor 's ($G_p(r)$). These models were fit using the same number of refinable parameters to allow for an unambiguous assessment of each model's validity. In the case of \Psmt , the starting crystallographic structure was expanded using the space group symmetry elements leading to a total of 16 atomic sites. Constraints were placed on the atomic positions and anisotropic ADP's such that the original symmetry operations were still enforced in the small box model. In addition to these parameters, a scale factor was refined and a term to account for correlated motion (see Ref.~\onlinecite{Farrow2007}).

The orthorhombic \Ammt\ model (see later sections for details of the distortion) results in an expansion of the unit to the new basis $(0,0,-1), (0,1,0), (2,1,0)$ in terms of the original lattice parameters $(a,a,c)$ and an increase in the number of symmetry independent atomic sites from 6 to 11. The lowering in symmetry from hexagonal to orthorhombic and the increased number of atomic sites creates additional refinable parameters in both the lattice parameters and the atomic positions. To account for this in our refinements, we constrained the orthorhombic \textit{a} and \textit{c} lattice parameters by a scale factor and refined the ADP's as isotropic. Furthermore, we constrained the ADP's for each atomic species to be equivalent (e.g. all As ADP's were refined by a single parameter). In doing so, the \Psmt\ model and the constrained \Ammt\ model were refined using the same number of parameters. We note that refinements of the \Ammt\ model were also performed without such constraints, the resulting fit produced lower goodness-of-fit parameters but qualitatively similar conclusions. We therefore, feel justified that the constraints are not artificially influencing the reported local structure.      

As discussed in the main text, the widely reported \Psmt\ model fits poorly in the low \textit{r} range - missing shoulders at \tild 2.6, and 5.4 \AA\ and systematically under/over-fitting adjacent peaks particularly of K-As character. Comparing the partial \Gor\ in panels \textit{a} and \textit{b} of Fig.~\ref{fig:sfour}, it is seen that the \Ammt\ model is able to account for this by creating a new series of Cr-As and K-As bond lengths - splitting the peak at 5.4 \AA\ (which is almost entirely K-As in character) and splitting the Cr-Cr peak at \tild 2.5 \AA . In fact, the \Gpor\ of As-As barely changes between the two. We attribute this distortion to a displacement of the K2 site and a redistribution of electron density in the Cr-Cr plaquettes (see main text).  

\begin{figure}[h]
	\includegraphics[scale=0.5]{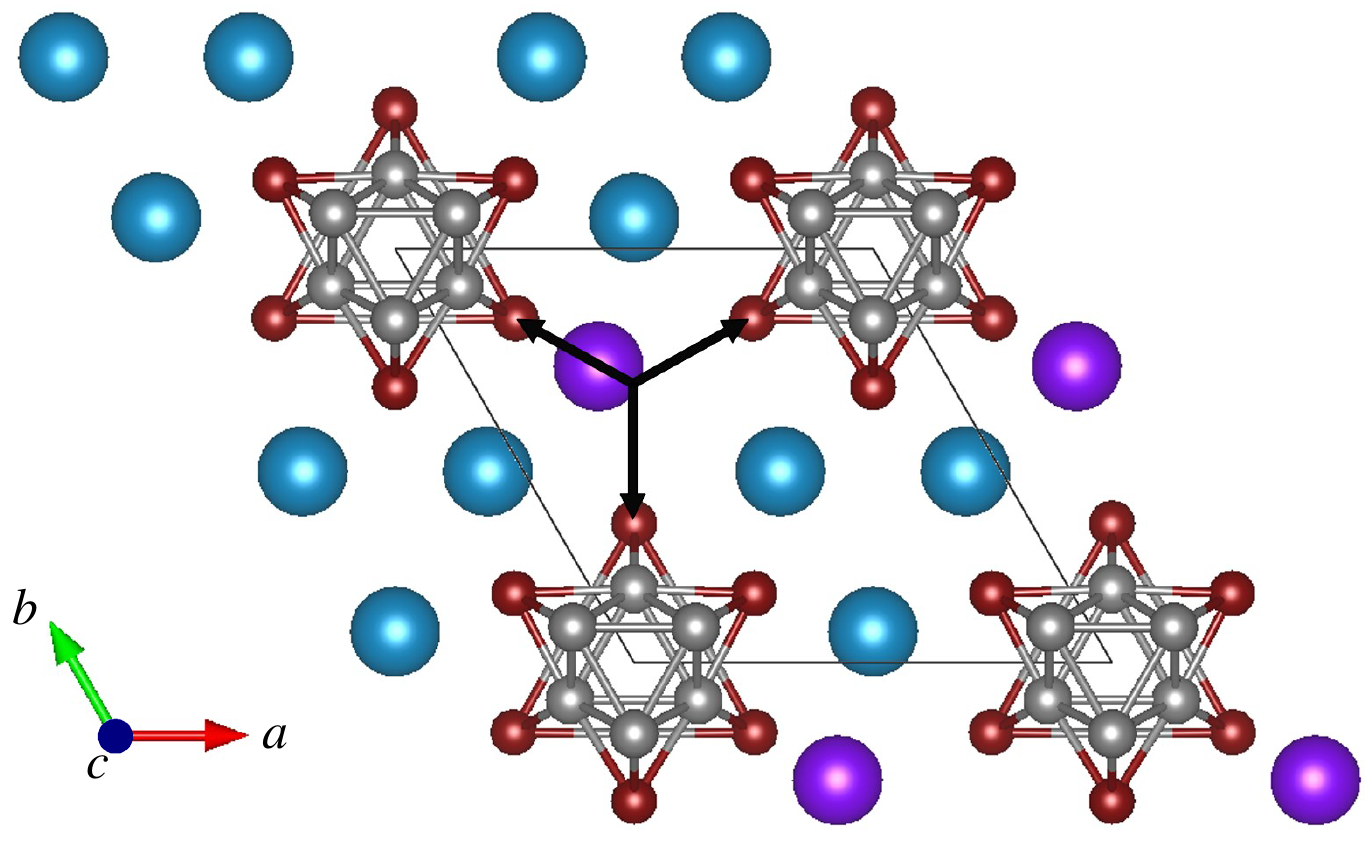}	
	\caption{\label{fig:sfive} Structure of \Psmt\ symmetry model refined with no constraints on in-plane K2 position. Black arrows added to indicate three equivalent displacement directions from crystallographic K2 site to neighboring CrAs tubes. The Cr, As, K1 and K2 sites are indicated by gray, red, teal and purple spheres respectively.}	
\end{figure}

To check this interpretation, refinements were performed with the \Psmt\ model but with the in-plane symmetry constraints of the K2 site's Wyckoff position removed (\textit{i.e.} $(x,y,z) = (\frac{1}{3}, \frac{2}{3},0)$). In contrast to the \Ammt\ model whose symmetry only allows K2 displacements along the (2,1,0) direction (in the original lattice setting), here the K2 is loosed to move any  in-plane direction in-plane. Refinements of this model result in reduced goodness-of-fit parameters (from $R_w$ \tild\ 14\% to 13\%) and a significant reduction of the K2 site's in-plane ADP's (by an order of magnitude form \tild\ 0.1 to 0.01 \Asq ). Considering the refined K2 position, it distorts from the special site to $(0.28, 0.71, 0)$ - along the line connecting the K2 and the CrAs tube center, in agreement with the \Ammt\ refinements (Fig.~\ref{fig:sfive}).

In addition to the $k = (0,0,0)$ \Ammt\ model as a simple check, small box modeling was also attempted using $k = (\frac{1}{2},0,0), (\frac{1}{2},\frac{1}{2},0)$ and $ (0,0,\frac{1}{2})$ distortions. We note that overall, these lower-symmetry models result in many additional refinable parameters. Especially considering the limited \textit{r} ranges used in our analysis, we treated additional refinable parameters with suspicion out of concern for artificially decreasing goodness-of-fit measures from increased model parameters and questions of over-fitting. Therefore, despite marginally better fits being achieved with distortions allowed by both attempted in-plane ordering vectors, we focus on the \Ammt\ model for its relative simplicity, the organic indication of a (2,1,0) displacement direction from the refinements described above and the results of our DFT calculations. On the other hand, models generated from the out of plane $ (0,0,\frac{1}{2})$ distortion direction were unable to improve our fit and were generally unstable as atoms were displaced from the $z = 0, \frac{1}{2}$ planes. This is unsurprising considering the results of our anisotropic ADP analysis which strongly indicate the distortion to be in-plane. 

\newpage

\section{\label{sec:symm} Distortions to the \Psmt\ structure }

The results of our DFT calculations, ADP analysis and PDF refinements in the \Psmt\ model with loosed K2 positions indicate the structural distortion to be in the \textit{ab} plane and likely along the K2 - Cr2 vector. Furthermore, the energy minimization work suggests a \Ammt\ orthorhombic distortion. Using these results as a starting point we performed Group Theory analysis of distortions to the \Psmt\ symmetry to further characterize the distortion. 


\begin{table*}[h]
	\caption{\label{tab:one}Irreducible representations ($\Gamma$), order parameter, space group and basis vectors for orderings with $k = (0,0,0)$ }
	\begin{ruledtabular}
		\begin{tabular}{lcccc}
     		 \multicolumn{1}{c}{$\Gamma$} & \multicolumn{1}{c}{Order parameter} & \multicolumn{1}{c}{Space group} & \multicolumn{1}{c}{Basis} \\
	\hline
	$GM_1 (k16t1)$ & $(a)$ &  $P\overline{6}m2$ &  $(1,0,0),(0,1,0),(0,0,1)$ \\
	$GM_2 (k16t3)$ & $(a)$ &  $P\overline{6}$ & $(1,0,0),(0,1,0),(0,0,1)$  \\
	$GM_3 (k16t4)$ & $(a)$ & $P3m1$ & $(1,0,0),(0,1,0),(0,0,1)$   \\
	\multirow{2}{*}{$GM_5 (k16t5)$} & $(a,-1.732a)$ & $Amm2$ & $(0,0,-1),(0,1,0),(2,1,0)$   \\
	                                & $(a,b)$ & $Pm$ & $(0,1,0),(0,0,1),(1,0,0)$   \\
	\multirow{3}{*}{$GM_6 (k16t6)$} & $(a,-1.732a)$ & $C2$ & $(0,-1,0),(2,1,0),(0,0,1)$ \\
									& $(a,0.577a)$ & $Cm$ & $(2,1,0),(0,-1,0),(0,0,-1)$ \\
									& $(a,b)$ & $P1$ & $(0,0,1),(1,0,0),(0,1,0)$ \\
		\end{tabular}
	\end{ruledtabular}
\end{table*}

Table~\ref{tab:one} shows $k = (0,0,0)$ allowed distortions. The corresponding irreducible representations (irreps $\Gamma$), order parameters, space groups and basis vectors were determined using the ISODISTORT program \cite{Campbell2006}. The $k = (0,0,0)$  produces five irreps and eight possible distorted structures. The first three irreps ($GM_1$, $GM_2$ and $GM_3$) maintain a screw/inversion symmetry element along the \textit{c}-lattice direction and so disallow in-plane displacements of the K2 site. The first irrep/order parameter combination to allow such a displacement is $GM_5/(a,-1.732a)$ which results in a new orthorhombic basis set and the \Ammt\ space group. The $E'(a)$ mode of $GM_5$ causes a displacement of the K2 site along the order parameter direction - along the (2,1,0) direction or \textit{c}-axis of the orthorhombic cell. This distortion is consistent with our PDF modeling in \Psmt\ with the in-plane symmetry constraints of K2 removed. As the highest symmetry model consistent with our previous modeling and with a symmetry predicted by our DFT work, we used the \Ammt\ symmetry for our small box modeling of the distortion. 

We note that while lower symmetry distortions with $k = (0,0,0)$ or with other in-plane ordering vectors also allowed such a distortion, they also resulted in a significant increase in the number of refinable parameters (see previous section) and therefore only report on the \Ammt\ model.

\newpage

\section{\label{DFT} Details of Density Functional Theory Calculations}

Density functional calculations were performed using the Perdew-Burke-Ernzerhof generalized gradient approximation (PBE-GGA)\cite{Perdew1996}. The lattice parameters were fixed at the experimental values, and all free internal atomic positions were relaxed subject to the hexagonal symmetry. As discussed, our results strongly differ from the previously reported stable phonons of Subedi. Accordingly, we did extensive convergence tests and employed two different electronic structure methods, specifically the projector-augmented wave method, as implemented in the VASP code Refs.~\onlinecite{Kresse1999,Kresse1996} and the general potential linearized augmented planewave method as implemented in WIEN2k Ref.~\onlinecite{Singh2006}. We carefully tested different convergence criteria, especially Brillouin zone samplings and basis set cut-offs in the VASP calculations. The phonons were obtained using the supercell finite difference method based on VASP calculations using the PHONOPY code \cite{Togo2015}. For this we tested different supercells. The result shown is for a $2\times 2\times 3$ supercell containing 192 atoms. As shown we find a strong phonon instability throughout the $k_z=0$ plane. We relaxed the structure, again using VASP, keeping the lattice parameters fixed, without imposing hexagonal symmetry and found a lower symmetry orthorhombic \Ammt\ structure. We did a similar relaxation using WIEN2k and also find that the hexagonal structure is unstable, and that the \Ammt\ structure is lower in energy similar to the VASP result. The calculated energy lowering is 17 meV per formula unit.

\begin{figure}[h]
	\includegraphics[scale=0.3]{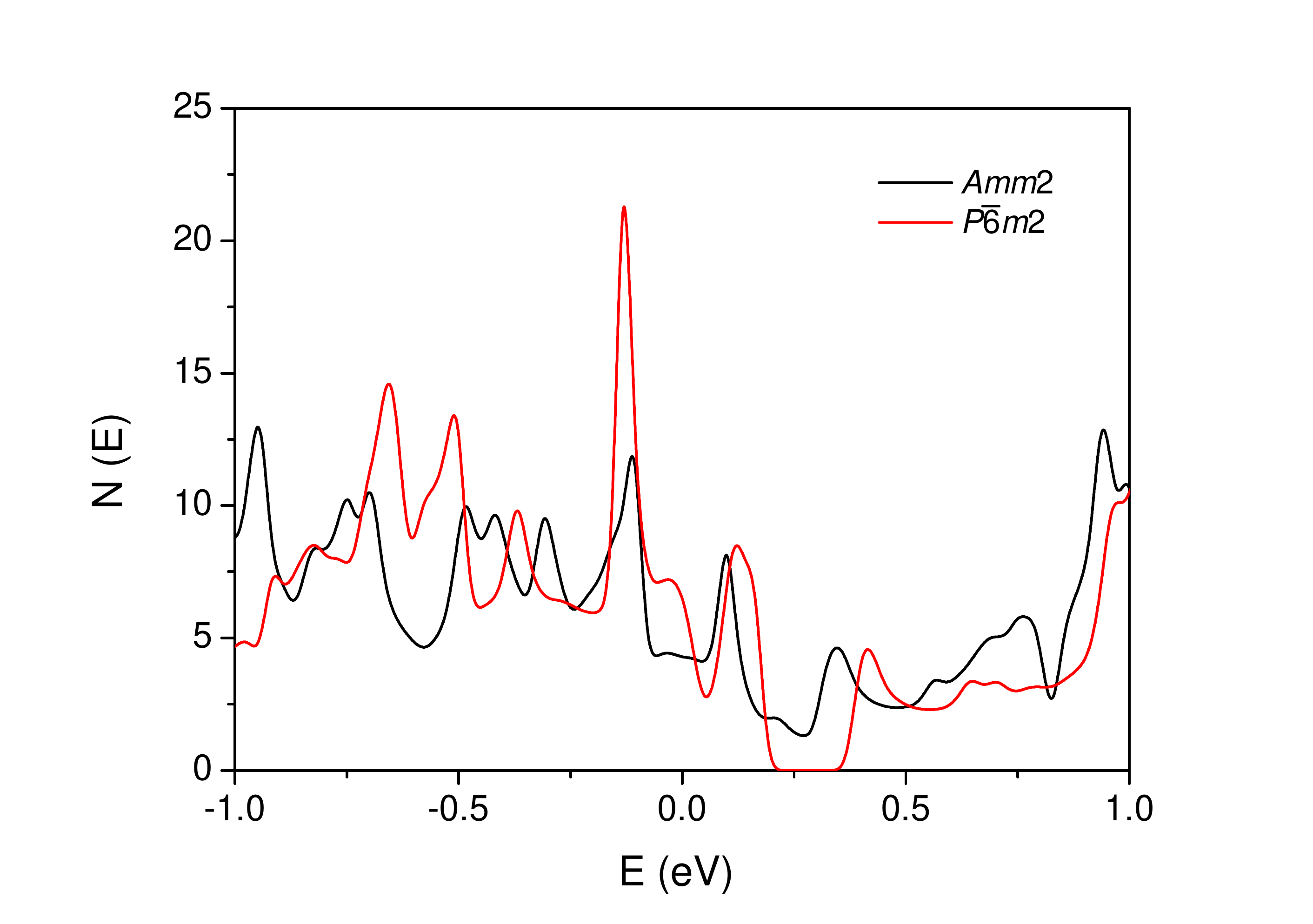}	
	\caption{\label{fig:ssix} Calculated density of states ($N(E)$) for the \Psmt\ and \Ammt\ structures in red and black respectively, with the Fermi energy ($E_{F}$) at $E = 0$ eV. }	
\end{figure}

To check for effects of the lowered \Ammt\ symmetry structure on transport properties, we calculated the electronic density of states ($N(E)$) for both the \Ammt\ and \Psmt\ symmetries using the VASP and WEIN2k methods (the results of the latter are shown in Fig.~\ref{fig:ssix}). Our calculations suggest a slight change in the shape of $N(E)$. However, this change does not significantly impact expectations of metallic transport behavior\cite{Cao2017}. In comparing these calculations to experiment it is important to note that the distortion we describe does not exhibit strong temperature dependence, does not have a transition temperature - being observed at all measured temperatures and is not long range. Therefore, previous transport measurements performed on the material ostensibly already include the effect of the local orthorhombic distortion. Furthermore, previous work comparing the calculated $N(E)$ to the experimental electronic specific-heat coefficients have suggested a significant electron mass renormalization, much larger than the difference we see here \cite{Cao2017}.  Finally, regarding the superconducting state, recent work (Ref.~\onlinecite{Liu2016}) has shown that the effect of non-magnetic disorder on \Tc\ in \ACA\ is reduced due to the relatively long mean free path and short coherence length of the superconducting pairs. This suggests the addition of such disorder might not have significant effect on the superconducting transition temperature.  These attributes complicate detection of the lowered symmetry structure in transport measurements.

\newpage